\documentclass[
aps, prl, reprint, superscriptaddress
]{revtex4-2}
\usepackage{amsmath,amssymb}
\usepackage{graphicx}
\usepackage{dcolumn}
\usepackage{bm}

\newcommand{\ket}[1]{\ensuremath{\left|#1\right>}}
\newcommand{\bra}[1]{\ensuremath{\left<#1\right|}}

\newcommand{\mf}[1]{\boldsymbol{#1}}
\newcommand{\mc}[1]{\ensuremath{\mathcal{#1}}}

\newcommand{\be}[1]{\begin{equation}#1\end{equation}}

\newcommand{\bal}[1]{\begin{align}#1\end{align}}
\newcommand{\bflal}[1]{\begin{flalign}#1\end{flalign}}

\begin{document}


\title{Electron-beam-induced quantum interference effects\\ in a multi-level quantum emitter}

\author{H. B. Crispin }
\email[Contact author: ]{crispin@physik.uni-kiel.de}
\affiliation{%
Institute for Experimental and Applied Physics, Christian Albrechts University, Leibnizstrasse 19, 24118 Kiel, Germany
}%
\author{N. Talebi}
\email[Contact author: ]{talebi@physik.uni-kiel.de}
\affiliation{%
Institute for Experimental and Applied Physics, Christian Albrechts University, Leibnizstrasse 19, 24118 Kiel, Germany
}%
\setcounter{footnote}{0}

\begin{abstract}
    Cathodoluminescence spectroscopy has recently emerged as a novel platform for nanoscale control of nonclassical features of light. 
    Here, we propose a theoretical model for cathodoluminescence from a multi-level quantum emitter. 
    Employing a master equation approach and treating the 
    electron-beam excitation as an incoherent broadband field source, 
    we show that quantum interference 
    can arise between the different relaxation pathways. 
    The induced-interference can significantly modify the time-dependent 
    spectra resulting in the 
    enhancement or suppression of cathodoluminescence.
    We find that the 
    excitation rate, initial state of the emitter, and excited level spacing
    play a crucial role in determining the influence of interference.   
    Our findings shed light on electron-beam-induced 
    quantum interference in 
    cathodoluminescence and 
    provides a theoretical basis 
    for exploring quantum optical phenomena in 
    electron-driven multi-level systems.    
\end{abstract}
\maketitle



Free-electron and matter interactions and their associated phenomena 
have attracted the attention of scientists for more than a century 
\cite{Talebi2017, *Polman2019, *Talebi2019a,cherenkov,*vavilov, 
ginfrank, *ginfrank2, Smith1953,Ritchie1957,clreview2017, clreviewnano}.
A well-known effect is cathodoluminescence (CL), in which
electrons incident on a material induce light emission \cite{clreview2017}. 
Recently, 
photon antibunching from nitrogen-vacancy 
(NV$^{0}$) centers
in  diamond nanoparticles has been observed while exciting 
them with electron beams \cite{Tizei2013}. This 
demonstrates the excitation of a two-level 
quantum emitter 
by fast electrons, thus paving the way for quantum optical studies with 
deep-subwavelength 
resolution. Subsequent works have shown the versatile nature of the 
CL 
photon statistics \cite{Meuret2015,
Meuret2017,Fiedler2023}. For instance, by varying the electron-beam 
current, a transition from sub-Poissonian to super-Poissonian 
behaviour was observed in the case of few emitters 
\cite{Meuret2015,Fiedler2023}. 
Moreover, superbunching effects emerge 
when multiple defect centers are excited, 
in contrast to the corresponding photoluminescence 
experiments \cite{Meuret2015}. The tunability of the photon statistics 
makes CL spectroscopy a novel technique 
for single-photon state generation, an essential requirement  
in quantum computing and quantum information 
\cite{sps, qcomp}. 
Therefore,  
a physical understanding of the nonclassical aspects of the 
CL emission is important 
from a theoretical and experimental perspective.

Several models have been proposed to understand 
the photon correlations in CL 
\cite{Meuret2015, cmodel,Feldman2018,Yanagimoto2021,mastereq}. 
The majority of these theoretical works, however, rely on 
classical models \cite{Meuret2015, cmodel,Feldman2018,Yanagimoto2021}.
Recently, an 
intuitive master equation approach was introduced to 
explain the superbunching effect \cite{mastereq}, where 
a phenomenological description of CL from 
two-level quantum 
emitters was provided.
The nature of the 
optical state of
the radiation field was also discussed based on heuristic arguments. 
However, 
 the focus has been 
on the electron-beam-excitations of only two-level systems so far. 
In a recent study
of the dephasing of hBN color centers \cite{Taleb2024}, a novel time-resolved CL spectroscopy 
technique  
has revealed the role of  
additional transition pathways in single emitters.
Therefore, exploring the dynamics of 
multi-level quantum emitters under electron-beam-excitations 
is already within experimental 
reach but a theoretical model for such systems 
is absent. 
\begin{figure*}[t]
    \centering
     \includegraphics[width=0.97\textwidth]{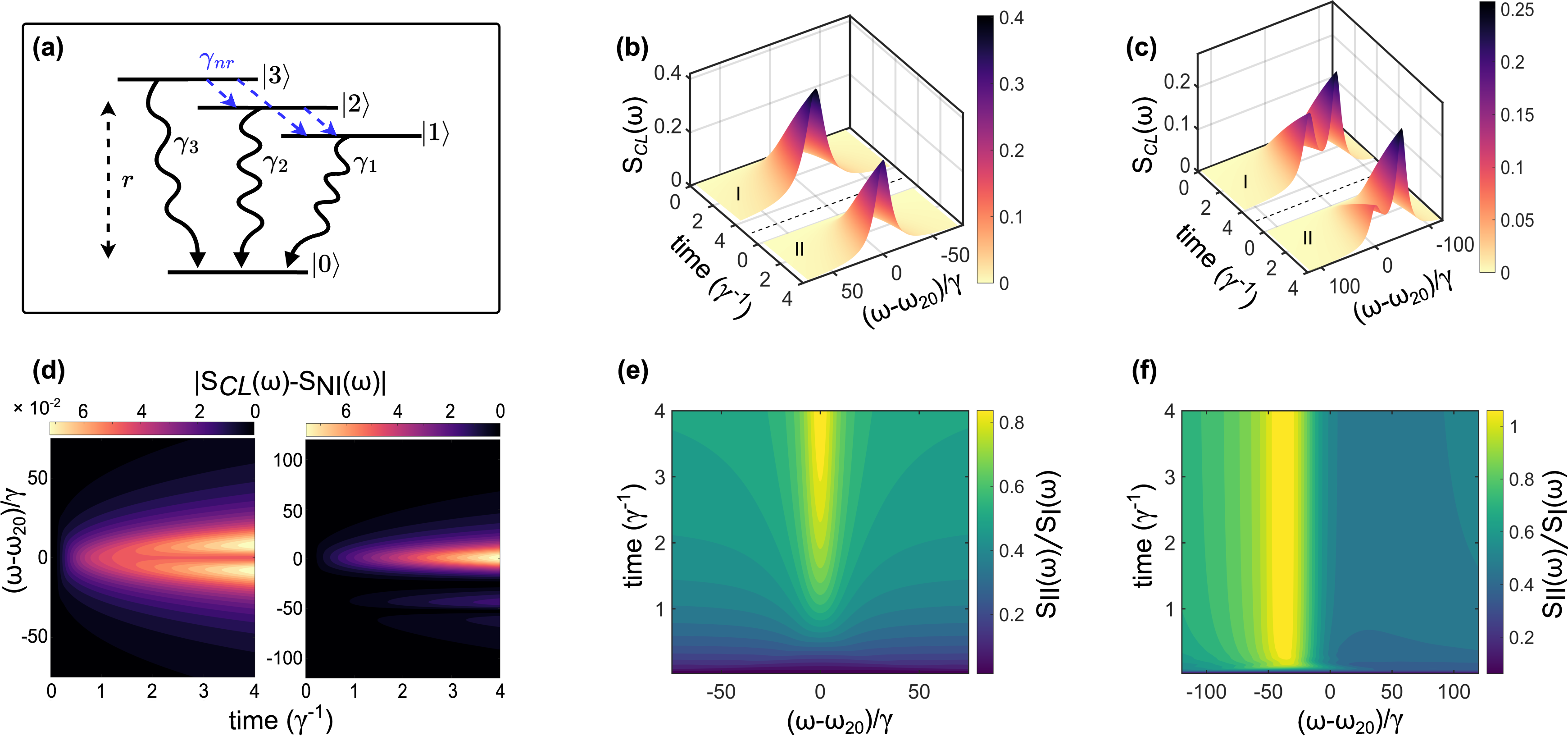}
\caption{(a) A four-level V-type configuration of the quantum emitter under consideration. 
        The energy levels $\ket{1}$, $\ket{2}$, and $\ket{3}$ 
        denote the excited states with decay rates $\gamma_{1}$, $\gamma_{2}$, 
        and $\gamma_{3}$, 
        respectively. 
        The states $\ket{3}$ and $\ket{2}$ 
        can also undergo nonradiative decay with rate $\gamma_{nr}$ to the 
        states $\ket{2}$ and $\ket{1}$, 
        as indicated by the blue arrows. 
        The electron-beam excitation modeled as an incoherent broadband field couples to 
        the transitions 
        $\ket{0}\longleftrightarrow\ket{1}$, 
        $\ket{0}\longleftrightarrow\ket{2}$, 
        and $\ket{0}\longleftrightarrow\ket{3}$ with $\mf{r}$ being the excitation rate. 
        In (b) and (c), 
        the time-dependent spectra is 
        plotted in the presence (curve II) 
        and absence (curve I) of interference, where the emitter initial state is 
        considered to be the ground state. 
        The considered parameters for simulations
         are $\gamma=1$, $r=5 \gamma$, $\gamma_{nr}=3\gamma$ for 
        two different excited level spacing of (b) $\omega_{21}=\omega_{32}=0.05 \gamma$ 
        and (c) $\omega_{21}=50\gamma, \omega_{32}=0.05 \gamma$. 
        (d) The interference contribution $(|S_{CL}(\omega)-S_{NI}(\omega)|)$ to the spectra as 
        a function of time and frequency for 
        $\omega_{21}=\omega_{32}=0.05 \gamma$ [left panel] 
        and $\omega_{21}=50\gamma, \omega_{32}=0.05 \gamma$ [right panel]. 
        The plots (e) and (f) show 
        the relative intensity of the spectral peaks 
        $(S_{\textrm{II}}(\omega)/S_{\textrm{I}}(\omega))$ associated with 
        the time-dependent spectra (a) and (b), respectively.
        }
\end{figure*}

In this work, we propose a theoretical framework for the 
electron-beam 
excitations of a single multi-level quantum emitter. 
We assume the system is continuously excited by incident electrons and 
undergoes radiative decay via spontaneous emission 
that is detected as CL. Our 
analysis relies on a quantum optical master 
equation that is derived from 
a model Hamiltonian describing 
the electron-emitter interaction. 
The resulting dynamical equations 
for the 
system naturally consist of the electron-beam-excitation part 
introduced phenomenologically
in Ref.~\cite{mastereq}. 
In addition, the broad spectral nature of the CL emission is predicted.
Moreover, a remarkable feature is the possibility of 
quantum interference between 
different transition pathways. 
The interference significantly alters the dynamics of the emitter and 
the time-dependent spectra leading 
to enhancement or suppression of spectral peaks. 
We find 
that the excitation rate, 
the initial state of the system, and 
the energy splitting between the excited levels 
play a crucial role in 
 determining the influence of induced interference. In general, 
 our master equation can be 
 applied to an $N$-level quantum emitter and the 
choice of the relevant energy levels, e.g., two levels or more, 
may depend on the 
material system and the transitions that are involved.

Here, we consider the emitter to be a four-level quantum 
system in a V-type configuration (see Fig. 1a). The excited states are denoted as 
$\ket{1}$, $\ket{2}$, and $\ket{3}$ that can spontaneously decay to 
the ground state $\ket{0}$ with rates $\gamma_{1}$, $\gamma_{2}$, and $\gamma_{3}$, 
respectively.
We also allow for nonradiative channels $\ket{3}\rightarrow\ket{1}$, 
$\ket{3}\rightarrow\ket{2}$, and $\ket{2}\rightarrow\ket{1}$ with $\gamma_{nr}$ being 
the nonradiative decay rate. This could represent
 the fast relaxation to the 
lower excited levels, e.g., via phonons.    
For simplicity, we assume this 
decay rate $\gamma_{nr}$ to be the same for all the nonradiative channels 
throughout our analysis. A typical 
quantum system with such specifications could be 
defects in two-dimensional materials coupled 
to nonradiative phonon excitations \cite{Zhang2024}. 
Since an incident electron 
excites optical
modes over a wide spectral range and CL at the nanoscale can be viewed as an 
analog of photoluminescence \cite{Tizei2013, Coenen2015}, we 
model the electron-emitter interaction by an incoherent broadband
pump field driving the quantum emitter.  
The influence of incoherent fields on atomic systems has been previously explored 
in a different context in quantum optics \cite{scully1,scully2,scully3,vic}. 
Here, we apply it 
to model the dynamics of the 
emitter under electron-beam-excitations. 
The density operator $\rho(t)$ of the system 
and its time evolution
in the Schr\"odinger picture (see the Supplementary Material
for the full derivation) is given by
\bal{\frac{d\rho}{dt}= &-\frac{i}{\hbar}[H_{0}, \rho]
                       -\!\sum_{i}(\gamma_{i}+\mf{r}_{i})
                       \left[\frac{1}{2}\{S^{+}_{i}S^{-}_{i},\rho\} - S^{-}_{i}\rho 
                       S^{+}_{i}\!\right]\nonumber\\
                       &-\sum_{i}\mf{r}_{i}\left(
                        \frac{1}{2}\{S^{-}_{i}S^{+}_{i},\rho\}
                       - S^{+}_{i}\rho S^{-}_{i}\right)\nonumber\\
                       &-\sum_{\substack{i,j\\i \ne j}}\mf{r}_{ij}
                       \bigg(\frac{1}{2}\{S^{+}_{i}S^{-}_{j}\!+\!S^{-}_{j}S^{+}_{i},\rho\}\!
                       - \!S^{-}_{j}\rho S^{+}_{i}\!
                       -\!S^{+}_{i}\rho S^{-}_{j}\!\!\bigg)\nonumber\\ 
                       &+\mc{L}_{nr}\rho. \label{qopeq}}   
where the system Hamiltonian $H_{0}=\sum_{n}\hbar \omega_{n0}\ket{n}\bra{n}$ ($n=1,2,3$), 
$\omega_{n0}=\omega_{n}-\omega_{0}$ is the transition frequency, $\gamma_{i}$ is the radiative 
decay rate of the $i^{th}$ transition, and 
$S^{+}_{i}\equiv 
A_{i0}=\ket{i}\bra{0}$ 
($S^{-}_{i}=(S^{+}_{i})^{\dagger}$, $i=1,2,3$) is the raising (lowering) operator for our system.   

The quantum optical master equation (\ref{qopeq}) models the 
dynamics of an electron-driven $N$-level quantum emitter. 
The complete derivation and the expression for the
nonradiative decay part ($\mc{L}_{nr}\rho$) is 
outlined in the Supplementary Material. 
The first and second terms in the master equation (\ref{qopeq}) 
describe the free evolution of the system 
and the familiar radiative 
decay of the excited levels \cite{vic,Scully1997}, 
respectively. 
More importantly, the third term in Eq.~(\ref{qopeq}) offers a key insight. 
It describes the 
electron-beam-excitation of the system (see Ref.~\cite{mastereq}), where the 
parameter 
$\mf{r}_{i}$ is 
the excitation rate of the emitter due to incident electrons.  
Here, the derivation of this excitation part naturally follows from
the interaction Hamiltonian (see Supplementary S1.).
Our focus throughout this Letter is the 
influence of the cross terms ($ i \ne j$) in Eq.~(\ref{qopeq}). 
 They indicate 
 the possibility of electron-beam-induced 
 quantum interference between different 
 transition pathways. 
 The interference amongst the $i^{\textrm{th}}$ 
 and $j^{\textrm{th}}$ ($i\ne j$) 
 transition 
 is denoted by the 
 cross term $\mf{r}_{ij}= p\sqrt{\mf{r}_{i}\mf{r}_{j}}$ \cite{scully3,vic}. Here, we 
 have introduced the interference parameter $p$. 
 $p=\pm 1$ 
 indicates maximal interference whereas $p=0$ implies no interference.   
 We show here that the presence of interference can greatly modify 
 the CL emission and the emitter dynamics. 
 Note that, in the simple case of a two-level system, quantum 
 interference is absent 
 and we obtain the phenomenological CL model \cite{mastereq} with
  $(\gamma+\mf{r})\rightarrow\gamma$. 
 
 To study the 
 effects of interference, we numerically compute the time-dependent CL spectrum of the 
 quantum emitter. Following Refs.~\cite{Eberly1977,Eberly1980}, the spectra for our system can be expressed in the form 
\bal{ S_{\textrm{CL}}(\omega)= ~&\textrm{Re}\int_{0}^{t}dt_{2}\int_{0}^{t-t_{2}}d\tau~
e^{-\Gamma(t-t_{2})} e^{(\Gamma/2-i\omega)\tau}\nonumber\\ &\times 
 \sum^{3}_{i,j=1} \gamma_{ij} \langle A_{i0}(t_{2}+\tau)A_{0j}(t_{2})\rangle,\label{spec}}
where $\Gamma$ is the filter bandwidth, 
$\omega$ is the observed frequency, 
$t$ is the elapsed time after the 
interaction, 
$\gamma_{ij}\equiv 2\gamma_{i}$ for $i=j$, 
 and 
$\gamma_{ij}\equiv-\sqrt{\gamma_{i}\gamma_{j}}$ for $i \ne j$. 
The derivation of the spectrum (\ref{spec}), evaluation of the 
two-time correlations, and the detection scheme are discussed 
in the Supplementary (see Sec. S2). 
In the following, 
we explore the influence of interference on the CL emission 
for different excited level 
separations, initial state conditions, and excitation rates 
using Eqs.~(\ref{qopeq})-(\ref{spec}). For the numerical analysis, we set   
$\gamma_{1}=\gamma_{2}=\gamma_{3}=\gamma$ and 
$\mf{r}_{1}=\mf{r}_{2}=\mf{r}_{3}=\mf{r}$, i.e., assuming 
equal decay and excitation rates. 
In all figures, the values of $\gamma_{nr}$ 
and $\Gamma$ are fixed as $\Gamma=0.1 \gamma$ and $\gamma_{nr}=3\gamma$ 
($\gamma_{nr}>\gamma$), respectively.  
We additionally scale all the parameters such as the excitation rate $\mf{r}$, 
nonradiative decay rate $\gamma_{nr}$, and 
frequencies ($\omega$, $\omega_{32}$, $\omega_{21}$) 
in units of $\gamma$. Unless stated otherwise, 
the initial state of the 
emitter is always considered to be the ground state.
 
First, we analyze the spectra (\ref{spec}) for different energy 
splitting of the excited levels. 
In Fig.~1(b), a 3d contour plot of the 
CL spectra is displayed 
for small 
energy splitting between the three 
upper levels 
($\omega_{21},\omega_{32}\ll \gamma$). To highlight the 
interference effects, 
we plot Eq.~(\ref{spec}) 
for the parameter values $p=0$ 
(without interference) and $p=1$ (with interference), 
indicated by 
I and II, respectively, in Fig.~1(b). 
The time-dependent spectra is single-peaked and  
centered around $\omega=\omega_{20}$, 
as expected for the near-degenerate case of 
the upper energy levels. We find that 
the quantum interference effect in  
the CL emission leads to the suppression of the central peak. 
This feature is more prominent for 
times $t\le\gamma^{-1}$ whereas at longer times $t>\gamma^{-1}$, the 
influence of interference diminishes. To see this temporal  
behaviour of the induced-interference, we plot the relative 
intensity of the spectral peaks of curves II and I in Fig.~1(e).
The contour clearly shows the time dynamics of the 
influence of interference 
in the CL emission. For times less than or comparable to the radiative 
lifetime of the excited-state levels ($t\le\gamma^{-1}$), 
the relative intensity around $\omega=\omega_{20}$ is $I<1$ (see Fig.~1(e)), 
depicted as shades of green in the plot. Therefore, the 
interference effects play a significant role at shorter time 
scales $t\le\gamma^{-1}$. 
In contrast, for longer times $t>\gamma^{-1}$, the relative intensity is
$I\approx1$, as can be seen from the transition from green to yellow
in the contour in Fig.~1(e). These results show the time-dependent 
behaviour of the induced-interference.

We consider now the case of large splitting 
between only two upper energy levels ($\omega_{21}\gg\gamma$) while 
the excited level $\ket{3}$ 
is near-degenerate with state $\ket{2}$ ($\omega_{32}\ll\gamma$). Fig.~1(c) 
shows the corresponding CL spectra. 
The emission at the central frequency stems 
from the decay of transitions $\ket{2}\rightarrow\ket{0}$ and 
$\ket{3}\rightarrow\ket{0}$ 
whereas the sideband 
originates from the transition $\ket{1}\rightarrow\ket{0}$. 
Interestingly, 
the strong suppression of the central peak due to interference is evident
for both $t\le\gamma$ and $t>\gamma$, 
in contrast to the near-degenerate case (see Fig.~1(c)). The contour plot 
in Fig.~1(f) shows the corresponding relative intensity 
of the curves II and I. The interference effects modify the 
emission at the central frequency $I<1$ 
whereas the sideband is almost 
unaffected $I\approx1$ (see Fig.~1(f)). Our analysis 
reveals that the 
quantum interference is not only 
sensitive to the level splitting but the time scale 
of its influence can significantly change accordingly. 
 
To gain more understanding of these results, we 
write the spectra as the sum of two parts 
$S_{CL}(\omega)=S_{I}(\omega)+S_{NI}(\omega)$. 
Here, $S_{I}(\omega)$ is the interference contribution 
to the total spectra and $S_{NI}(\omega)$ is referred to as
 the noninterfering part that can be obtained by setting $p=0$ 
in Eq.~(\ref{spec}). It is useful to visualize only 
the interference contribution by plotting the quantity 
$|S_{CL}(\omega)-S_{NI}(\omega)|$. The corresponding contour plot
is shown in Fig.~1(d) for the two cases of excited level splitting 
discussed in Figs.~1(b)-(c). 
For closely spaced upper energy levels (left panel 
in Fig.~1(d)), 
note that the interference contribution is  
at the central frequency only upto times 
$t=2\gamma^{-1}$ but shifts to $\omega=\omega_{20}\pm\omega_{s}$ 
for $t>2\gamma^{-1}$. In contrast, for the 
nondegenerate case, we have the maximum interference 
contribution always centered around $\omega=\omega_{20}$, 
as shown in Fig.~1(d) (right panel). 
 \begin{figure*}[t]
    \centering 
    \includegraphics[width=0.99\textwidth, height=3.7cm]{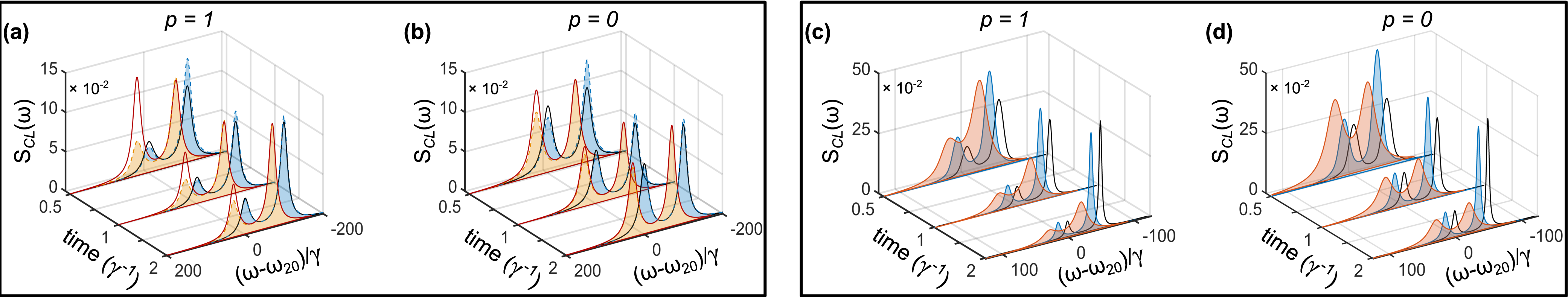}
     \caption{The evolution of the spectra for different initial states of the emitter is plotted in (a) [with interference, $p=1$] and (b) 
     [without interference, $p=0$] at times $t=0.5$, $t=1$, and $t=2$ (units of $\gamma^{-1}$). The parameters are $\gamma=1$, $r=5 \gamma$, 
     and $\omega_{21}=100\gamma,\,\omega_{32}=0.05\gamma$.
     The solid black and red curves 
     show the spectra computed with the initial state as the ground state $\ket{\psi(t_{0})}=\ket{0}$ and an equally weighted 
     superposition state $\ket{\psi(t_{0})}=\sum^{3}_{i=0}c_{i}e^{i \delta_{i}}\ket{i}$ 
      ($c_{i}=1/2,\,\delta_{12}=0,\delta_{13}=\delta_{23}=\pi$), respectively. The blue and orange shaded 
      curves show the spectra for the 
      initial states $\ket{\psi(t_{0})}=[\ket{0}+\ket{1}]/\sqrt{2}$ and $\ket{\psi(t_{0})}=\sum^{3}_{i=0}c_{i}e^{i \delta_{i}}\ket{i}$ 
      ($c_{i}=1/2,\,\delta_{i}=0$), respectively. For clarity, the orange (shaded) and red (solid) curves are shifted by 30 units along the $\omega$-axis, 
      and the spectra at times $t=0.5$ and $t=1$ 
      are multiplied by a factor of $4$ and $1.5$, respectively. The dependence on the excitation rate in (c) the presence and (d) absence of 
      interference and the spectra at times $t=0.5$, $t=1$, and $t=2$ (units of $\gamma^{-1}$). 
      The parameters are $\mf{r}=0.5 \gamma$ (solid black), $\mf{r}=\gamma$ (shaded blue), $\mf{r}=5\gamma$ (shaded red), and  $\omega_{21}=50\gamma,\,\omega_{32}=0.05\gamma$ with 
       the initial state as the ground state. For clarity, the blue and red curves (shaded) are shifted by 15 and 45 units along the $\omega$-axis, 
       respectively and the spectra at times $t=0.5$ and $t=1$ 
       are multiplied by a factor of $15$ and $3$, respectively.}
   \end{figure*}
   
   We now proceed to study the spectra for various 
   initial conditions.
    Specifically, we are interested 
   in the role of interference when the emitter is 
   prepared in a different quantum state prior to the interaction. 
   This is displayed in 
   Figs.~2(a)-(b). In Fig.~2(a), the time-dependent 
   spectra for four different initial states of  
   the quantum emitter is plotted, 
   taking into account the 
   effects of interference ($p=1$). 
   Fig.~2(b) shows the respective 
   CL spectra in the absence of quantum interference ($p=0$). 
   Comparing Fig.~2(a) and 2(b), the 
   spectra strongly depends on the initial state of the 
   system when considering interference effects. At short timescales 
   $t\le\gamma^{-1}$, we find that the spectra changes drastically 
   if the emitter is prepared in an initial coherent superposition state
   rather than the ground state (compare the colored curves to the 
   solid black curve in Fig.~2(a)).  
   Tuning the relative phase of the initial superposition allows us to control 
   the quantum interference effect. This is shown in Fig.~2(a) 
   (compare orange curve to solid red curve for time $t=0.5\gamma^{-1}$), where 
   the emission from the upper levels ($\ket{2}, 
   \ket{3}$$\rightarrow\ket{0}$) is significantly enhanced,  
   even in the presence of nonradiative decay. 
   Although these 
   modifications are less prominent for $t=2\gamma^{-1}$, 
   a notable difference between the spectra for the ground state 
   and that of the superposition state (compare solid black and red curves
    in Fig.~2(a)) is still evident. On the contrary, the CL emission is almost 
   independent of the initial state in the absence of interference 
   (compare solid black and red curves
    in Fig.~2(b) for $t=2\gamma^{-1}$). We may thus consider the initial-state dependence of the 
    spectra as a strong signature of
    the electron-beam-induced interference in time-resolved CL.
  
    \begin{figure}[t]
    \centering 
    \includegraphics[scale=0.25]{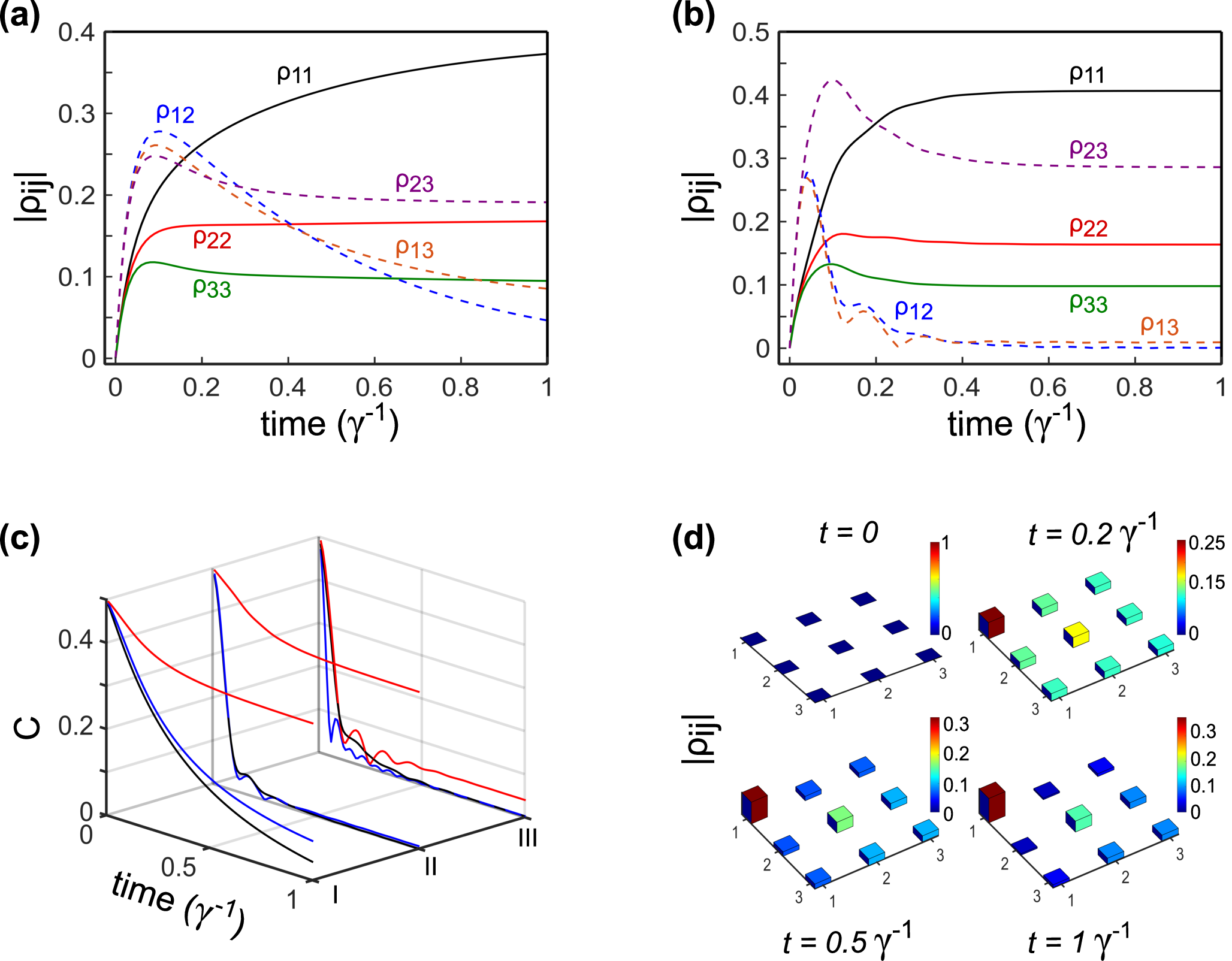}
       \caption{The excited-state populations and the coherence dynamics plotted in (a) and (b). The parameters are $\gamma=1$, $r=5 \gamma$,
          (a) $\omega_{21}=\omega_{32}=0.05 \gamma$ 
        and (b) $\omega_{21}=50\gamma, \omega_{32}=0.05 \gamma$. The initial state is considered to be the ground state in all subplots. For clarity, the coherences (dashed curves)
         in (a) and (b)  
        are multiplied by a factor of 2 and 3, respectively.
        The $3$d plot in (c) shows the time evolution of the 
        ratio $C=|\rho_{ij}|/(\rho_{ii}+\rho_{jj})$ ($i \ne j$) 
        of 
        the two-photon coherences to the populations for different excited level separations: (I) $\omega_{21}=\omega_{32}=0.05 \gamma$, 
        (II) $\omega_{21}=50\gamma, \omega_{32}=0.05 \gamma$, and (III) $\omega_{21}=\omega_{32}=50 \gamma$. The solid black, blue, and red curves 
        show the parameter $C$ computed for two-photon coherences $\rho_{12}$, $\rho_{13}$, and $\rho_{23}$, respectively. (d) The magnitude of 
        the density matrix elements $|\rho_{ij}|$ ($i,j=1,2,3$) at times $t=0$, $t=0.2$, $t=0.5$, and $t=\gamma$ with the parameters being the same as in (a).}
    \end{figure}

    An important parameter in CL experiments that can be controlled is the 
    electron-beam current $I$. One can relate it  
    to the excitation rate $\mf{r}$ \cite{mastereq}. We now consider different 
    excitation rates and analyze the 
    CL emission. Figs.~2(c)-(d) display 
    the spectra for weak 
    ($\mf{r}\le\gamma$) and strong excitations ($\mf{r}>\gamma$). 
   Following a similar analysis like in our previous discussion, 
   we plot the spectra for both the interference parameter values 
   $p=1$ (Fig.~2(c)) and $p=0$ (Fig.~2(d)). A common characteristic 
   feature seen in both scenarios 
   is the broadening of the spectral peaks. For larger values of the 
   excitation rate, we find that the effect is stronger (compare red curve 
   to solid black curve in Figs.~2(c)-(d)). This behavior can be easily 
   understood from the
    master equation (\ref{qopeq}). It can be clearly seen that 
    the radiative decay width of the 
   excited levels is the sum of the 
   natural line width and the excitation rate $\gamma_{i}+\mf{r}_{i}$. 
   The electron-beam-excitation 
   therefore leads to broad spectral 
   features in the CL emission. 
   
   
  Next, we investigate the corresponding density 
  matrix dynamics of the emitter.
  In Fig.~3(a)-(b), we plot the populations and the two-photon coherences 
  for the near-degenerate and nondegenerate cases 
  explored earlier in Figs.~1(b)-(c). A 3d barchart 
  of the density matrix elements and its time evolution 
  is shown in Fig.~3(d) considering the same parameters as in Fig.~3(a).  
  The interference leads to population transfer between the 
  excited levels, as seen from nonzero values of 
  two-photon coherences $|\rho_{ij}|$ ($i,j=1,2,3$) in 
  Figs.~3(a)-(b) (dashed curves). The two-photon coherences grow 
  from zero to a maximum value before starting to 
  decay and reaching a steady-state value. On shorter time scales 
  $t<1/\mf{r}$, there can be strong interference even for 
  large splitting between the two excited levels (see Fig.~3(b)). 
  To better understand the dynamics of the quantum path interferences, 
  we compute the ratio $C=|\rho_{ij}|/(\rho_{ii}+\rho_{jj})$. Fig.~3(c) 
  shows the parameter $C$ for three different energy splitting of 
  the excited-state levels. In the situation where atleast two 
  upper levels are near-degenerate (curves I and II in Fig.~3(c)), 
  the interference effects do play an important role, as 
  the parameter $C$ saturates at a steady-state (red curve in I and II) leading 
  to observable effects in the spectra (Fig.~1(b)-(c)). 
  However, for a large splitting between all the excited levels, 
  the ratio $C$ rapidly decays to zero (curve III in Fig.~3(c)). 
  The interference effects  
  are therefore negligible 
  in such cases. 

  We now discuss the possible 
  experimental scenarios and parameter regimes to test the predictions. 
  In CL, the 
  excitation process usually involves additional elementary processes 
  mediated by secondary electrons or phonons \cite{Meuret2017,prawer2014}. The timescale 
  associated with the excitation of the 
  emitter is very small, ranging from 
  femtoseconds to picoseconds. In comparison to the radiative 
  lifetime of the emitter that is of the order of nanoseconds, the excitation 
  timescale is much shorter, i.e., $t_{exc}\ll t_{rad}$. Therefore the 
  relevant parameter regime to explore 
  would be $\mf{r}\gg \gamma$, where $\mf{r}$ is the 
  excitation rate and $\gamma$ 
  represents the radiative decay rate of the emitter. 
  Note that this parameter $\mf{r}$ may be controlled via  
  the electron beam current $I$ 
  in experiments. 
  For the experimental realization,
  the time-resolved
   CL spectroscopic technique 
  based on electron-driven photon sources would be a suitable choice 
  \cite{Talebi2019,Nielen2020,Christopher2020,Taleb2023}.
   Particularly, 
  the spectra for different initial conditions can be 
  explored using this setup. 
  This is realized by a tailored metamaterial for 
  illuminating the sample with focussed coherent radiation.
  The system can therefore be excited to an initial coherent 
  superposition state prior to the electron-beam-excitations. 
  In addition, to probe the dynamics of a single emitter, 
  subnanometer resolution is required, 
  as demonstrated in 
  Ref.~\cite{Tizei2013}. 
  The investigation of quantum interference phenomena 
  is therefore feasible with 
  current CL spectroscopic techniques.  

  In conclusion, we have developed a theoretical framework for 
  cathodoluminescence from a quantum emitter. Modeling the 
  electron-emitter interaction by an 
  incoherent broadband field driving the system, we derive a 
  quantum optical master equation for an electron-driven multi-level 
  emitter. The master equation accounts for the electron-beam-excitation 
  of the emitter and the broad spectral nature of CL light emission.
  The present work therefore provides a new insight into 
  the electron-matter interactions in CL and goes beyond 
  previous phenomenological approaches. 
  A 
  novel result is 
  the possibility of electron-beam-induced quantum interference between 
  different transition pathways. We find that induced interference  
  leads to significant modifications of the spectra and the emitter dynamics. 
  Depending on the initial state preparation and the splitting between 
  the excited levels, the quantum interference 
  may either enhance or suppress the CL emission. These results 
  can be crucial for 
  time-resolved measurements of CL at the nanometerscale.  
   Our findings open up the exploration of 
   electron-beam-induced interference phenomena and provides a 
   theoretical basis for investigating 
   quantum optical effects in spectroscopic CL. 
   Further generalization of this scheme, considering 
   multiple emitters, paves the way towards exploring many-body 
   correlations and superradiance features.

   \textit{Acknowledgments:} This project has received funding from the Volkswagen Foundation
(Momentum Grant), European Research Council (ERC) under the European
Union’s Horizon 2020 research and innovation program under grant
agreement no. 802130 (Kiel, NanoBeam), and from Deutsche
Forschungsgemeinschaft.

   \bibliographystyle{apsrev4-2} 
  \clearpage 
  \onecolumngrid
  \section*{Supplementary Material: 
\textbf{Electron-beam-induced quantum interference effects 
in a multi-level quantum emitter}}

\subsection*{\normalsize \textbf{S1.} 
\textbf{Derivation: 
Master equation for a multi-level quantum 
emitter}}
In this section, we derive the incoherent broadband 
field contribution to 
the master equation (Eq.~(1) in the main text). 
The theory describing the spontaneous 
radiative decay of the excited levels is well-established and 
can be found in standard   
quantum optics 
textbooks and monographs [1, 2]. 
Here, we focus
on the electron-beam-induced dynamics of 
the system. 
We model the CL interaction between the incident electrons and 
the emitter by a  
broadband incoherent pump field exciting the multi-level quantum emitter. 
The corresponding electric field can be written as 
\bflal{
    \vec{E}(t)=\mc{E}_{e}(t)e^{-i \omega_{e}t}\hat{e}+\textrm{c.c.},
    }
where $\mc{E}_{e}(t)$ is the field amplitude, $\omega_{e}$ is the 
central frequency, and $\hat{e}$ denotes the polarization vector. 
Since the field is incoherent,
the amplitude $\mc{E}_{e}(t)$ is assumed to be $\delta$-correlated at 
different times and 
having zero mean value, 
\be{
    \langle \mc{E}_{e}(t)\mc{E}^{*}_{e}(t')\rangle=\mc{R} \, \delta(t-t')~~~~
    \langle \mc{E}_{e}(t)\rangle=0
}
Here, $\mc{R}$ denotes the incoherent excitation strength. 
The interaction Hamiltonian 
for our system reads
\bal{
    H_{I}= -\sum^{3}_{i=1}\mc{E}_{e}(t)\big(\vec{d}_{i0}\cdot \hat{e}\,  
    A_{i0}\big)
   e^{-i \omega_{e}t}
    +\textrm{H.c.,}
}
where $\vec{d}_{i0}$ ($i=1,2,3$) are the transition dipole moments and  
$A_{i0}=\ket{i}\bra{0}$ ($i=1,2,3$)
denote the transition operators. 
In this work, it is assumed that 
the dipole moments are real and mutually orthogonal.  
The equations of motion 
for the combined state of the emitter and the field is given by the 
Liouville-Von Neumann equation
\be{
    \frac{\partial \tilde{\rho}(t)}{\partial t}=-\frac{i}{\hbar}
    [\tilde{H}_{I}(t), \tilde{\rho}(t)].\label{dme}
}  
In the above equation, 
the density operator $\tilde{\rho}(t)=\textrm{U}(t)\rho(t)\textrm{U}^{\dagger}(t)$ 
and the 
Hamiltonian $\tilde{H}_{I}(t)=\textrm{U}(t)H_{I}(t)\textrm{U}^{\dagger}(t)$ are
expressed in 
the interaction picture according to  
$\textrm{U}(t)=\exp\{i H_{0} t /\hbar\}$, where  
$H_{0}=\sum_{i}\hbar \omega_{i0}\ket{i}\bra{i}$ 
($i=1,2,3$) is the system Hamiltonian. A formal integration of Eq.~(\ref{dme}) leads to 
the solution
\be{
    \tilde{\rho}(t)=\tilde{\rho}(t_{0})-\frac{i}{\hbar}\int_{t_{0}}^{t} dt'\,
    [\tilde{H}_{I}(t'),\tilde{\rho}(t')].
} 
Substituting the above expression back in Eq.~(\ref{dme}) results in 
an integro-differential 
equation of the form 
\be{\frac{\partial \tilde{\rho}(t)}{\partial t}=-\frac{i}{\hbar}
 [\tilde{H}_{I}(t), \tilde{\rho}(t_{0})]-\frac{1}{\hbar^{2}}
 \int_{t_{0}}^{t} dt'\,
 [\tilde{H}_{I}(t),[\tilde{H}_{I}(t'),\tilde{\rho}(t')]].
}
Since we are mainly interested in the dynamics of the emitter rather 
than the combined quantum 
state of the emitter and the field, we perform a partial trace 
over the variables of the subsystem corresponding to 
the incoherent field. The time evolution of the reduced density operator 
is then given by
\be{\frac{\partial \tilde{\rho}_{qe}(t)}{\partial t}=-\frac{i}{\hbar}
 \textrm{Tr}\,[\tilde{H}_{I}(t), \tilde{\rho}(t_{0})]-\frac{1}{\hbar^{2}}
 \textrm{Tr}\,\int_{t_{0}}^{t} dt'\,
 [\tilde{H}_{I}(t),[\tilde{H}_{I}(t'),\tilde{\rho}(t')]],
}
where $\tilde{\rho}_{qe}(t)=\textrm{Tr}\{\tilde{\rho}(t)\}$ is the state 
of the emitter. Within the framework of the 
generalized reservoir theory and the Born approximation [3, 4], 
the incoherent broadband field is 
considered to be a large reservoir characterized 
by infinite degrees of freedom 
that is weakly coupled to the emitter. Therefore, the 
 dynamics of the emitter is significantly 
modified by its interaction with the broadband field whereas the 
the influence of the emitter on the field
 is assumed to be negligible. We can thus approximate
 the total density operator as 
 $\tilde{\rho}(t)=\tilde{\rho}_{qe}(t)\otimes \rho_{p}(t_{0})$ at all times 
 [3], where 
 $\rho_{p}(t_{0})$ is the state of the broadband field reservoir. With these 
 assumptions and making use of Eqs.~(2)-(3) in Eq.~(7), we obtain the following 
 equation for the density operator $\tilde{\rho}_{qe}(t)$,
 \bal{
    \frac{\partial \tilde{\rho}_{qe}(t)}{\partial t}=
    &-\frac{1}{2} \sum_{i,j}\bigg[
\frac{(\hat{d}_{i0}\cdot \hat{e})(\hat{d}_{j0}\cdot 
\hat{e})^{*}\mc{R}}{\hbar^{2}}(S^{+}_{i}S^{-}_{j}\tilde{\rho}_{qe}+
\tilde{\rho}_{qe} S^{+}_{i}S^{-}_{j} - 2S^{-}_{j}\tilde{\rho}_{qe} 
S^{+}_{i})e^{i\omega_{ij}t}\nonumber\\
&~~~~~~~~~~+\frac{(\hat{d}_{i0}\cdot \hat{e})^{*}(\hat{d}_{j0}\cdot 
\hat{e})\mc{R}}{\hbar^{2}}(S^{-}_{i}S^{+}_{j}\tilde{\rho}_{qe}+
\tilde{\rho}_{qe} S^{-}_{i}S^{+}_{j} - 2S^{+}_{j}\tilde{\rho}_{qe} 
S^{-}_{i})e^{-i\omega_{ij}t}\bigg].
     } 
Dropping the subscripts and undoing the unitary 
transformation $\textrm{U}=\exp\{i H_{0} t /\hbar\}$, we can write 
the equation for the density operator in the Schrodinger picture as 
\bal{\frac{\partial \rho}{\partial t}= &-\frac{i}{\hbar}[H_{0}, \rho]
                       -\sum_{i}\mf{r}_{i}
                       \left[\frac{1}{2}\{S^{+}_{i}S^{-}_{i},\rho\} - S^{-}_{i}\rho 
                       S^{+}_{i}\right]
                       -\sum_{i}\mf{r}_{i}\left(
                        \frac{1}{2}\{S^{-}_{i}S^{+}_{i},\rho\}
                       - S^{+}_{i}\rho S^{-}_{i}\right)\nonumber\\
                       &-\sum_{\substack{i,j\\i \ne j}}\mf{r}_{ij}
                       \bigg(\frac{1}{2}\{S^{+}_{i}S^{-}_{j}+S^{-}_{j}S^{+}_{i},\rho\}
                       - S^{-}_{j}\rho S^{+}_{i}
                       -S^{+}_{i}\rho S^{-}_{j}\bigg). \label{meq}}
The above expression gives the incoherent broadband field contribution to the
 master equation of the system with 
$\mf{r}_{ij}= p_{ij}\sqrt{\mf{r}_{i}\mf{r}_{j}}$ and 
excitation rate $\mf{r}_{i} = |\vec{d}_{i0}\cdot \hat{e}|^{2}\mc{R}/\hbar^2$, where 
$p_{ij}=(\vec{d}_{i0} \cdot \hat{e})(\vec{d}_{j0} \cdot \hat{e})^{*}
/(|\vec{d}_{i0} \cdot \hat{e}||\vec{d}_{j0} \cdot \hat{e}|)$ is the 
introduced interference parameter [4]. For our 
calculations, we 
consider $|\vec{d}_{10}|=|\vec{d}_{20}|=|\vec{d}_{30}|=|d|$ and 
$\mf{r}_{1}=\mf{r}_{2}=\mf{r}_{3}=\mf{r}$. Therefore, as mentioned 
in the main text, 
we have $p_{ij}=p$ ($i,j=1,2,3$, $i \ne j$).
The second term in Eq.~(\ref{meq}) is responsible for the radiative broadening 
while the third term describes the 
electron-beam-excitation of the emitter. 
The cross terms $i\ne j$ lead to quantum interference effects. 
We incorporate the nonradiative 
decay of the excited levels ($\ket{3}\rightarrow\ket{1}$, 
$\ket{3}\rightarrow\ket{2}$, and $\ket{2}\rightarrow\ket{1}$) 
phenomenologically into the master equation according to the
Lindblad operator 
\bal{\mc{L}_{nr}\rho=&-\gamma_{nr}\left[
    \frac{1}{2}\{A_{22}, \rho\}-A_{12}\rho A_{21}\right]
    -\gamma_{nr} \bigg[\{A_{33}, \rho\}-A_{13}\rho A_{31}-A_{23}\rho A_{32} \bigg],
     } 
where the nonradiative decay rate $\gamma_{nr}\gg \gamma$.
The final master equation used in our analysis is therefore given by   
\bal{\frac{\partial \rho}{\partial t}= &-\frac{i}{\hbar}[H_{0}, \rho]
                       -\sum_{i}(\gamma_{i}+\mf{r}_{i})
                       \left[\frac{1}{2}\{S^{+}_{i}S^{-}_{i},\rho\} - S^{-}_{i}\rho 
                       S^{+}_{i}\right]
                       -\sum_{i}\mf{r}_{i}\left(
                        \frac{1}{2}\{S^{-}_{i}S^{+}_{i},\rho\}
                       - S^{+}_{i}\rho S^{-}_{i}\right)\nonumber\\
                       &-\sum_{\substack{i,j\\i \ne j}}\mf{r}_{ij}
                       \bigg(\frac{1}{2}\{S^{+}_{i}S^{-}_{j}+
                        S^{-}_{j}S^{+}_{i},\rho\}
                       -S^{-}_{j}\rho S^{+}_{i}
                       -S^{+}_{i}\rho S^{-}_{j}\bigg)+\mc{L}_{nr}\rho. }   
In the presence of interference, one finds that the resulting density matrix equations for 
the populations and two-photon coherences are coupled, thereby 
leading to interesting effects in 
the dynamics and the CL spectra, as shown 
in Figs.~(1)-(3). To solve for the excited-state populations and the
coherence dynamics of the emitter, we define a column vector of the 
one-time averages
\bal{\Psi(t)=&(\langle A_{11} \rangle, \langle A_{22} \rangle, 
\langle A_{33} \rangle, \langle A_{12} \rangle, \langle A_{21} \rangle, 
\langle A_{13} \rangle, 
\langle A_{31} \rangle
\langle A_{23} \rangle, \langle A_{32} \rangle, \nonumber\\&
\langle A_{10} \rangle, 
\langle A_{01} \rangle, \langle A_{20} \rangle, \langle A_{02} \rangle, 
\langle A_{30} \rangle, \langle A_{03} \rangle, \langle A_{00} \rangle)^{T},}
where $\rho_{ij}=\langle A_{ji} \rangle$. The column vector $\Psi(t)$ satisfies 
the equation 
\be{
    \frac{d \Psi(t)}{dt}=M \Psi(t).
}
Here, $M$ is a $16\times16$ square matrix whose elements are coefficients
that appear in the density matrix equations (11). It is now convenient to solve 
the matrix differential 
equation (13) numerically and find the time evolution of 
the density matrix elements. In addition, we can 
use it to evaluate the two-time 
correlation functions in the  
time-dependent spectra (see Sec. \textbf{S2}).

\subsection*{\normalsize \textbf{S2.} 
\textbf{Time-dependent spectra and evaluation of the two-time correlations}}
A rigorous theoretical description of the 
time-dependent spectrum was given by Eberly and 
W\'odkiewicz [5]. This 
generalization of the steady-state 
spectra takes into account the filtering of the light field, e.g., by 
a Fabry-Perot interferometer, prior to detection. 
The physical spectrum depends on 
the first-order correlation functions of the electric-field operator 
and has been studied in the 
context of spontaneous emission [5, 6, 7] and 
resonance fluorescence from two-level and three-level systems [5, 8, 9]. More 
recently, the Eberly-W\'odkiewicz spectrum 
was used to investigate the time dynamics of the  
photoluminescence spectra 
of a hBN color center [10]. 
In general, the time-dependent spectrum can be written as
\be{
    S(\omega, \Gamma; t)=
    \Gamma \int_{t_{0}}^{t}\!\!dt_{1}\int_{t_{0}}^{t}\!\!dt_{2}\,\,\,
    e^{-(\Gamma/2-i\omega)(t-t_{1})}
    e^{-(\Gamma/2+i\omega)(t-t_{2})}\,
    \langle \hat{\vec{E}}^{(-)}(\vec{r}, t_{1})\cdot 
    \hat{\vec{E}}^{(+)}(\vec{r}, t_{2})\rangle, \label{tspec}
}
where $\omega$ is the observed frequency, $\Gamma$ is the spectral bandwidth of 
the filter, and $t_{0}=0$ is considered the initial time. For the four-level V-type system 
under consideration, 
the negative frequency part of 
the electric-field operator [11] at the position $\vec{r}=r \hat{r}$ 
of the detector in the far-field reads
\be{
    \hat{\vec{E}}^{(-)}(\vec{r}, t)=-\frac{\eta}{c^{2}r}\sum_{i=1}^{3}
    \omega_{i}^{2}\{\hat{r}\times(\hat{r}\times \vec{d}_{i})\}
    \,S^{+}_{i}(t-r/c)\,e^{ikr},
} 
where $\eta=1/4\pi\epsilon_{0}$, $\omega_{i}\equiv \omega_{i0}$ ($i=1,2,3$) 
denotes the transition frequency, 
$\vec{d}_{i}\equiv \vec{d}_{i0}$ ($i=1,2,3$) 
is the transition dipole moment, and 
$S^{+}_{i}\equiv A_{i0}=\ket{i}\bra{0}$ denotes the raising operators. 
We consider a general
detection 
scheme in which all the cross-correlations $\langle A_{i0}(t_{2}+\tau)
A_{0j}(t_{2})\rangle$ ($i \ne j$) also contribute to the 
emission spectra 
(\ref{tspec}) by choosing 
the unit vector $\hat{r}$
in the direction of observation 
to be $\hat{r} \cdot \vec{d}_{i} = |d|\cos\alpha=|d|/\sqrt{3}$, 
where $|\vec{d}_{10}|=|\vec{d}_{20}|=|\vec{d}_{30}|=|d|$. 
Substituting now the expression for the 
electric field operator (15) 
in the spectra (14) 
and making a 
change of variables $t_{1}\rightarrow t_{2}+\tau$ [5], we find
\bal{
  S(\omega, \Gamma)=~&\phi \,\,
  \textrm{Re}\int_{0}^{t}dt_{2}\int_{0}^{t-t_{2}}d\tau~
  e^{-\Gamma(t-t_{2})} e^{(\Gamma/2-i\omega)\tau}\nonumber\\& \times
   \bigg(\sum^{3}_{i=1} 2\gamma 
   \langle A_{i0}(t_{2}+\tau)A_{0i}(t_{2})\rangle
   -\gamma \sum^{3}_{\substack{i,j=1\\i \ne j}} 
   \langle A_{i0}(t_{2}+\tau)A_{0j}(t_{2})\rangle \bigg),  
}
where we have assumed $\omega_{i0}/\omega_{j0}\approx 1$ 
($i \ne j$) in arriving at the above equation 
with a prefactor
$\phi=\Gamma \hbar \omega_{10} /(2 c r^{2})$ and 
 the decay rate defined as 
 $\gamma=|d|^{2}\omega_{10}^{3}/(3\pi\epsilon_{0}\hbar c^{3})$. 
 We scale $S(\omega, \Gamma)$ with respect to $\phi$ and express 
 the time-dependent CL spectra as 
\bal{S_{CL}(\omega, t)\equiv S(\omega, \Gamma; t)/\phi=~&
\textrm{Re}\int_{0}^{t}dt_{2}\int_{0}^{t-t_{2}}d\tau~
e^{-\Gamma(t-t_{2})} e^{(\Gamma/2-i\omega)\tau}
\sum^{3}_{i,j=1} \gamma_{ij} \langle A_{i0}(t_{2}+\tau)A_{0j}(t_{2})\rangle,
}
where the two-time correlations are written in a compact form by defining  
$\gamma_{ij}\equiv 2\gamma_{i}$ for $i=j$ 
and 
$\gamma_{ij}\equiv-\sqrt{\gamma_{i}\gamma_{j}}$ for $i \ne j$. 
Eq.~(17) gives the form of the time-dependent CL spectra used in the main text.

To calculate the two-time correlation functions in Eq.~(17), we introduce a 
column vector of the two-time averages given by
\bal{
    \vec{Y}^{mn}(t_{2}, \tau)=
    &\big[Y_{11}(t_{2}, \tau),Y_{22}(t_{2}, \tau), 
    Y_{33}(t_{2}, \tau), 
    Y_{12}(t_{2}, \tau), 
    Y_{21}(t_{2}, \tau), \nonumber\\&
    Y_{13}(t_{2}, \tau), 
    Y_{31}(t_{2}, \tau), 
    Y_{23}(t_{2}, \tau),
    Y_{32}(t_{2}, \tau), 
    Y_{10}(t_{2}, \tau), \nonumber\\&
    Y_{01}(t_{2}, \tau), 
    Y_{20}(t_{2}, \tau), Y_{02}(t_{2}, \tau)
    Y_{30}(t_{2}, \tau), Y_{03}(t_{2}, \tau), Y_{00}(t_{2}, \tau)\big]^{T}, ~~~~m,n=0,1,2,3
}  
with 
\be{
    Y_{ij}(t_{2}, \tau)=\langle A_{ij}(t_{2}+\tau) A_{mn}(t_{2}) \rangle.
}
Applying the quantum regression theorem [5,12] results in 
\be{
    \frac{d\, \vec{Y}^{mn}(t_{2}, \tau)}{d\tau}=M\, \vec{Y}^{mn}(t_{2}, 0). 
} 
The solution can be expressed as 
$\vec{Y}(t_{2}, \tau)=e^{M \tau} \vec{Y}(t_{2}, 0)$, where $\vec{Y}(t_{2}, 0)$ 
is determined from the differential equations governing the one-time averages (13). 
Following Ref.~[5], we can 
introduce a matrix $\mc{T}^{mn}$ such that 
$\vec{Y}^{mn}(t_{2}, 0)=\mc{T}^{mn} \Psi(t_{2})$, which allows us to conveniently evaluate 
the integral in Eq.~(17) given the initial conditions $\Psi(t_{0})$.

 \section*{ References}
\noindent $[1]$ M. O. Scully and M. S. Zubairy, Quantum Optics 
    (Cambridge University Press, 1997).\\
    $[2]$ Z. Ficek and S. Swain,~Quantum interference and 
    coherence: Theory and Experiments 
    (Springer-Verlag, 2005).\\
    $[3]$ V. Kozlov, Y. Rostovtsev and M. Scully, 
     Phys. Rev. A. \textbf{74}, 063829 (2006).\\
    $[4]$ M. Kiffner, M. Macovei, J. Evers and C. H. Keitel, 
     Prog. Opt. \textbf{55}, 85-197 (2010).\\
     $[5]$ J. H. Eberly, C. V. Kunasz, and K. W\'odkiewicz, J. Phys.
     B: At. Mol. Phys. \textbf{13}, 217 (1980).\\
     $[6]$ J. E. Golub and T. W. Mossberg, 
     Phys. Rev. Lett. \textbf{59}, 2149 (1987).\\
     $[7]$ J. J. Sanchez-Mondragon, N. B. Narozhny, 
     and J. H. Eberly,
     Phys. Rev. Lett. \textbf{51}, 550 (1983); \textbf{51} 1925 (1983).
    $[8]$ A. S. Jayarao, S. V. Lawande, and R. D’Souza, 
     Phys. Rev. A \textbf{39}, 3464 (1989).\\
     $[9]$ R. Román-Ancheyta, O. de los Santos-Sánchez,
      L. Horvath, and H. M. Castro-Beltrán, 
      Phys. Rev. A \textbf{98}, 013820 (2018). \\
      $[10]$ D. Groll, T. Hahn, 
      P. Machnikowski, D. Wigger, and T. Kuhn, 
      Mat. Quant. Tech. \textbf{1}, 015004 (2021).\\
      $[11]$ G. S. Agarwal and S. Menon, 
      Phys. Rev. A \textbf{63}, 023818 (2001).\\
      $[12]$ M. Lax, Phys. Rev. \textbf{129}, 2342 (1963).

\end{document}